\documentclass[conference,a4paper]{IEEEtran}
\IEEEoverridecommandlockouts

\usepackage[IEEEtran,notheorems,url,graphics,hyperrefblack]{research17}
\usepackage{cite}
\usepackage{balance}
\usepackage{booktabs} 
\usepackage{enumitem}
\usepackage{mathtools}
\usepackage[lined,boxed,commentsnumbered,linesnumbered,ruled]{algorithm2e}
\setlist[description]{leftmargin=\parindent,labelindent=\parindent}
\usepackage{tikz} 
\usetikzlibrary{calc} 
\newcommand{\tikzmark}[1]{\tikz[overlay,remember picture] \node (#1){};}
\newtheorem{lemma}{\mylemmaname}
\newtheorem{theorem}{\mytheoremname}
\newtheorem{definition}{\mydefinitionname}
\newtheorem{proposition}{\mypropositionname}
\newtheorem{corollary}{\mycorollaryname}
\newtheorem{example}{\myexamplename}
\usepackage{multirow}
\usepackage{array}
\newcolumntype{C}[1]{>{\centering\arraybackslash}p{#1}}
\usepackage{amsmath}
\makeatletter
\renewcommand*\env@matrix[1][*\c@MaxMatrixCols c]{%
  \hskip -\arraycolsep
  \let\@ifnextchar\new@ifnextchar
  \array{#1}}
\makeatother
\usetikzlibrary{decorations.pathreplacing}
\usepackage{twoopt} 
\newcommandtwoopt{\tikzoverbrace}[5][raise=3mm,amplitude=3pt][yshift=4mm]{\draw
  [decorate,decoration={brace,#1}] (#3) -- (#4) node [font=\small, above, pos=.5,#2] {#5};}
\newcommandtwoopt{\tikzunderbrace}[5][raise=1mm,amplitude=3pt][yshift=-1mm]{\draw 
  [thick,decorate,decoration={brace, mirror,#1}] (#3) -- (#4) node [font=\small, below, pos=.5,#2] {#5};}
\allowdisplaybreaks
\newcommand{\collect}[1]{\mathscr{#1}} 
\newcommand{\prop}[1]{\mathsf{#1}} 
\begin{document}
\title{Lengthening and Extending Binary\\ Private
  Information Retrieval Codes}

\author{
  \IEEEauthorblockN{Hsuan-Yin Lin and Eirik Rosnes}
  \IEEEauthorblockA{Simula@UiB, N--5020 Bergen, Norway\\
  (Emails: hsuan-yin.lin@ieee.org and eirikrosnes@simula.no)
  \thanks{This work was partially funded by the Research Council of Norway (grant 240985/F20).}}
}

\maketitle

\begin{abstract}
  It was recently shown by Fazeli \emph{et al.}  that the storage
  overhead of a traditional $t$-server private information retrieval
  (PIR) protocol can be significantly reduced using the concept of a
  \emph{$t$-server PIR code}. In this work, we show that a family of
  $t$-server PIR codes (with increasing dimensions and blocklengths)
  can be constructed from an existing $t$-server PIR code through
  lengthening by a single information symbol and code extension by at
  most $\bigl\lceil t/2\bigr\rceil$ code symbols. Furthermore, by
  extending a code construction notion from Steiner systems by Fazeli
  \emph{et al.}, we obtain a specific family of $t$-server PIR
  codes. Based on a code construction technique that lengthens and
  extends a $t$-server PIR code simultaneously, a basic algorithm to
  find good (i.e., small blocklength) $t$-server PIR codes is
  proposed. For the special case of $t=5$, we find provably optimal
  PIR codes for code dimensions $k \leq 6$, while for all
  $7 \leq k \leq 32$ we find codes of smaller blocklength than the
  best known codes from the literature. Furthermore, in the case of
  $t=8$, we also find better codes for $k=5,6,11,12$. Numerical
  results show that most of the best found $5$-server PIR codes can be
  constructed from the proposed family of codes connected to Steiner
  systems.
\end{abstract}


\section{Introduction}
\label{sec:introduction}

Private information retrieval (PIR) has attracted significant attention for well over a decade since its introduction by
Chor \emph{et al.} in \cite{ChorKushilevitzGoldreichSudan95_1}.  A formal PIR protocol allows to privately retrieve a
single file among the servers storing it without revealing any information about the requested file to each individual
server. Traditional PIR protocols operate on a database of $n$ bits, which is replicated among several servers to achieve
PIR. Thus, the storage overhead of traditional PIR protocols is at least $2$, and the overall goal is to reduce the
total upload and download cost of the protocol.

PIR for distributed storage systems was first addressed in \cite{ShahRashmiRamchandran14_1}. 
For distributed storage systems the size of the requested file is typically much larger
than the number of files, and thus the upload cost is much lower than the download cost. Hence, only the download cost
is considered, as opposed to traditional PIR protocols. Recent work on PIR protocols for distributed storage systems
typically assumes that the storage code is given, and then the PIR protocol is designed as a second layer to the system
\cite{TajeddineRouayheb16_1,KumarRosnesGraell17_1}. This is in contrast to the work by Fazeli \emph{et al.} in
\cite{FazeilVardyYaakobi15_1sub}, where, in order to reduce the storage overhead of traditional PIR protocols, the
concept of a \emph{$t$-server PIR code} was proposed. A $t$-server PIR code is an $[n,k]$ linear code satisfying the
so-called $t$-PIR property, i.e., for every information symbol, there exist $t$ mutually disjoint subsets of
$\{1,2,\ldots,n\}$ such that it can be recovered from the code symbols indexed by any of these $t$ subsets. By employing an
$[n,k]$ $t$-server PIR code, they have shown that all known $t$-server information-theoretic PIR protocols can be
emulated by a coded PIR protocol with storage overhead equal to $n/k$. 

Finding good codes that operate efficiently with a small storage overhead, i.e., designing a $t$-server PIR code
with a small blocklength for a given dimension, is an important research challenge. In \cite{FazeilVardyYaakobi15_1sub}, an insightful series
of $t$-server PIR code constructions based on existing code construction techniques were presented. In the recent work
of \cite{VajhaRamkumarKumar17_1}, the authors found that the so-called \emph{shortened projective Reed Muller (SPRM)}
codes are good $t$-server PIR codes for $t=2^\ell-1$ and $2^\ell$ where $\ell$ is a positive integer. For $t=3,4$, it was
shown in \cite{VajhaRamkumarKumar17_1} that SPRM codes are indeed optimal in the sense of achieving a lower bound on the
blocklength of a $t$-server PIR code.

In this work, we will show that a $t$-server PIR code with small blocklength can be constructed by lengthening and
extending an existing PIR code. Furthermore, we prove that a certain family of codes associated with Steiner systems
possesses the $t$-PIR property. Since optimal codes for $t\leq 4$ are known (see
\cite{FazeilVardyYaakobi15_1sub,VajhaRamkumarKumar17_1}), we mainly focus on the special case of $t=5$ (or,
equivalently, $t=6$) for which we show that provably optimal PIR codes can be constructed from lengthening and extending
an existing PIR code for code dimensions $k \leq 6$, while for all $7 \leq k \leq 32$ we find codes of smaller
blocklength than the best known codes from the literature. Moreover, we also show that for certain values of $k$, SPRM
codes are not optimal for $t=8$.


\vspace*{-0.25mm}
\section{Definitions and Preliminaries}
\label{sec:definitions-preliminaries}

Throughout this paper, we will focus on binary codes only.
Component-wise addition of vectors from a vector space will be written as normal addition, and as is customary in coding
theory, we denote row vectors by boldface italic Roman letters, e.g., $\vectg{x}$. However,
sometimes we will slightly abuse this notational convention by using $\vectg{c}$ to refer to a column vector. Moreover,
whether an all-zero vector $\vectg{0}$ (or an all-one vector $\vectg{1}$) is a row vector or a column vector
will become clear from the context. The Hamming weight of a binary vector $\vectg{x}$ is denoted by
$w_\textnormal{H}(\vectg{x})$ throughout the paper.

\subsection{$t$-Server PIR Codes}
\label{sec:PIRcodes_t}

\begin{definition}
  \label{def:PIRcodes_t}
  Consider an $[n,k]$ linear code $\code{C}$ and its corresponding
  generator matrix
  $\mat{G}\eqdef\bigl[\vectg{c}_1,\ldots,\vectg{c}_n\bigr]$. This
  $[n,k]$ code is said to be an $[n,k;t]$ PIR code if for every
  $i\in\Naturals_k\eqdef\{1,2,\ldots,k\}$, there exist $t$ mutually
  disjoint sets $\set{R}_h^{(i)}$, $h\in\Naturals_t$, such that
  \begin{IEEEeqnarray*}{rCl}
    \vectg{e}_i& \eqdef &\trans{(\underbrace{0,\ldots,0}_{i-1},1,0,\ldots,0)}=\sum_{j\in\set{R}^{(i)}_h}\vectg{c}_j,
    \quad\forall\;h\in\Naturals_t,
  \end{IEEEeqnarray*}
  where superscript ``$\trans{}$" denotes vector transposition. Equivalently, if we define the codeword
  $\vectg{x}=\vectg{u}\mat{G}$ to be the encoding of $\vectg{u}$, for every $u_i$, $i\in\Naturals_k$, we have
  \begin{IEEEeqnarray*}{rCl}
    u_i=\sum_{j\in\set{R}^{(i)}_h} x_j,\quad\forall\;h\in\Naturals_t.
  \end{IEEEeqnarray*}
  We also say that such a code $\code{C}$ (or $\mat{G}$) has the $t$-PIR property. Moreover, given a message symbol
  $u_i$, $i\in\Naturals_k$, those mutually disjoint sets $\set{R}_h^{(i)}$, $h\in\Naturals_t$, are called the recovering
  sets for $u_i$.
\end{definition}

For given values of $k$ and $t$, the minimum value of $n$ for which an $[n,k;t]$ PIR code exists is of great
interest. This motivates us to look at a related parameter in conventional coding theory: the length of the shortest
binary linear code with dimension $k$ and minimum Hamming distance $d$. The smallest blocklength of a linear code for
fixed values of $(k,d)$ has been discussed extensively in the existing literature (see, e.g.,
\cite{MacWilliamsSloane77_1}). Note that our notation of an $[n,k;t]$ PIR code should not be confused with the usual
three parameters notation of an $[n,k,d]$ linear code, where the third parameter $d$ denotes the minimum Hamming
distance of the $[n,k]$ code. We make the following definitions.
\begin{definition}
  \begin{IEEEeqnarray*}{rCl}
    N_\textnormal{P}(k,t)& \eqdef &\min\{n\colon \textnormal{an }
    [n,k;t] \textnormal{ binary PIR code exists}\}.
    \\
    N(k,d)& \eqdef &\min\{n\colon \textnormal{an } [n,k,d]
    \textnormal{ binary linear code exists}\}.
  \end{IEEEeqnarray*}
\end{definition}

\subsection{Bounds for $t$-Server PIR Codes}
\label{sec:bounds-PIRcode_t}

It is well-known that the minimum Hamming distance $d$ of a $t$-server
PIR code must be at least $t$ \cite{vitaly17_1}.
\begin{proposition}
  \label{prop:d_tPIRcodes}
  If an $[n,k;t]$ PIR code exists, then its minimum Hamming distance $d$
  must satisfy $d\geq t$.
\end{proposition}

\begin{corollary}
  \label{cor:lower-bound_tPIRcodes}
  For given values of $k$ and $t$, $N_\textnormal{P}(k,t)$ is lower-bounded by the smallest blocklength $n$ such that an
  $[n,k,t]$ code exists, i.e., $N_\textnormal{P}(k,t)\geq N(k,t)$.
\end{corollary}
\begin{IEEEproof}
  We prove the inequality by contradiction. Suppose there exists an $[n,k;t]$ PIR code such that $n<N(k,t)$. Then, by
  definition this implies that the minimum Hamming distance of such a PIR code is less than $t$, which leads to a
  contradiction to Proposition~\ref{prop:d_tPIRcodes}.
\end{IEEEproof}

In \cite{VajhaRamkumarKumar17_1}, a lower bound on the minimum blocklength $N_\textnormal{P}(k,t)$ for any
\emph{systematic} $[n,k;t]$ PIR code was presented. As shown in \cite{RaoVardy17_1sub}, the bound from
\cite{VajhaRamkumarKumar17_1} also holds for any binary $[n,k;t]$ PIR code. The lower bound from
\cite{VajhaRamkumarKumar17_1}, denoted by $L_\textnormal{P}(k,t)$, is
\begin{IEEEeqnarray*}{rCl}
  L_\textnormal{P}(k,t)\eqdef k+\left\lceil\sqrt{2k+\frac{1}{4}}+\frac{1}{2}\right\rceil+t-3,\quad t\geq 3.
\end{IEEEeqnarray*}
It can easily be verified that in general $N(k,t)\geq L_\textnormal{P}(k,t)$ for small values of $t>4$. In fact, we will
show in Section~\ref{sec:numerical-results} that $N(k,t)$ is a tighter lower bound on $N_\textnormal{P}(k,t)$ than
$L_\textnormal{P}(k,t)$ for $t=6$.

Some useful upper and lower bounds on $N_\textnormal{P}(k,t)$ were provided by Fazeli \emph{et al.}  in
\cite{FazeilVardyYaakobi15_1sub}. Together with the constructions introduced therein, the authors provided an upper
bound table on $N_\textnormal{P}(k,t)$ for all values of $k\leq 32$ and $t\leq 16$. We briefly summarize their results
below.
\begin{lemma}[Lemmas 13 and 14 in \cite{FazeilVardyYaakobi15_1sub}]\leavevmode
  \label{lem:bounds_tPIRcodes}
  \begin{enumerate}[label={(\alph*)}]
  \item \label{item:a} $N_\textnormal{P}(k,t+t') \leq N_\textnormal{P}(k,t)+N_\textnormal{P}(k,t')$,
  \item \label{item:b} $N_\textnormal{P}(k+k',t) \leq N_\textnormal{P}(k,t)+N_\textnormal{P}(k',t)$,
  \item \label{item:c} $N_\textnormal{P}(k,t)\leq N_\textnormal{P}(k+1,t)-1$,
  \item \label{item:d} $N_\textnormal{P}(k,t)\leq N_\textnormal{P}(k,t+1)-1$, and
  \item \label{item:e} if $t$ is odd, then $N_\textnormal{P}(k,t+1) = N_\textnormal{P}(k,t)+1$.
  \end{enumerate}
\end{lemma}

\section{Code Constructions}
\label{sec:code-constructions}

In this section, we first present a code construction by lengthening and extending a given PIR code, and then present an
extension of a code construction inspired by Steiner systems proposed by Fazeli \emph{et al.} in
\cite{FazeilVardyYaakobi15_1sub}. An earlier work constructing PIR codes (and even stronger
batch codes) for $t=k$
based on Steiner systems (and more general block designs) was presented in \cite{WangShakedCassutoBruck13_1}.

\subsection{Lengthening and Extending PIR Codes}
\label{sec:lengthening-extending_tPIRcodes}

In the following theorem, we will investigate an important property of a PIR code with an arbitrary positive integer
$t$.

\begin{theorem}
  \label{thm:lengthen-tPIRcode}
  For any given $t\in\Naturals\eqdef\{1,2,\ldots\}$, we have
  \begin{IEEEeqnarray*}{rCl}
    N_\textnormal{P}(k+1,t)\leq
    N_\textnormal{P}(k,t)+\left\lceil\frac{t}{2}\right\rceil.
  \end{IEEEeqnarray*}
\end{theorem}
\begin{IEEEproof}
  The proof is based on the fact that when $t$ is even, we can always construct a new generator matrix of an
  $[n+\frac{t}{2},k;t+1]$ code by adding one length-$n$ row of weight $\frac{t}{2}$ and appending the column vector $\vect{e}_{k+1}$ a total of $\frac{t}{2}$ times to the existing generator matrix of an $[n,k;t]$ code. See the detailed derivations in
  Appendix~\ref{sec:proof_lengthen-tPIRcode}.
\end{IEEEproof}

Theorem~\ref{thm:lengthen-tPIRcode} is an improved version of part~\ref{item:b} of Lemma~\ref{lem:bounds_tPIRcodes} for
$k'=1$, while for $k'>1$, it is an improved version only if
$k'\left\lceil\frac{t}{2}\right\rceil<N_\text{P}(k',t)$. This theorem suggests that for a given even value of $t$, a new
$t$-server PIR code can always be generated by adding one information symbol and appending at most $t/2$ code symbols to
the original $t$-server PIR code.

Next, we will discuss a special family of systematic codes that will help in the numerical search for good PIR codes
with small blocklength, especially when $k$ is large.

\subsection{Construction of PIR Codes Based on Steiner Systems}
\label{sec:PIRcodes_Steiner}

In \cite{FazeilVardyYaakobi15_1sub}, a systematic code construction based on Steiner systems was proposed, in which the
authors introduce a representation method of systematic codes, and give a sufficient (but not necessary) condition for
constructing PIR codes.

\begin{definition}
  \label{def:representation_systematic-code}
  Let $\collect{P}_k=\{\set{P}_j\}_{j=1}^r$ be a collection of subsets of $\Naturals_k$. A systematic $[n=k+r,k]$ code
  $\code{C}$ can be represented by defining the codewords of $\code{C}$ as
  $\vectg{x}\eqdef(u_1,\ldots,u_k,x_{k+1},\ldots,x_{k+r})$, where $u_1,\ldots,u_k$ are the information bits of the code
  and each redundancy bit $x_{k+j}$ is defined as $x_{k+j}\eqdef\sum_{i\in\set{P}_j}u_i$, $j\in\Naturals_r$.
 
  We denote the constructed code by $\code{C}(\collect{P}_k)$. Furthermore, for the sake of notational convenience, we
  define $\set{J}^{(i)}\eqdef\bigl\{j\in\Naturals_r\colon i\in\set{P}_j\bigr\}$ to be the set of indices
  $j\in\Naturals_r$ such that $i\in\set{P}_j$.
  
  The systematic generator matrix $\mat{G}$ of this code can be written as
  $\mat{G} = \bigl[\mat{I}_k | \mat{P}_{k\times r}\bigr]$, where $\mat{I}_k$ is the $k\times k$ identity matrix and the
  $k\times r$ redundancy matrix $\mat{P}_{k\times r}=\{p_{i j}\}_{1\leq i\leq k,\,1\leq j\leq r}$ is defined by
  \begin{IEEEeqnarray*}{rCl}
    p_{i j}\eqdef
    \begin{cases}
      1, & \textnormal{if } i\in\set{P}_j,
      \\
      0, & \textnormal{otherwise}.
    \end{cases}
  \end{IEEEeqnarray*}
\end{definition}

\begin{lemma}[Lemma 7 in \cite{FazeilVardyYaakobi15_1sub}]
  \label{lem:tPIRcode_Steiner}
  Suppose that a collection $\collect{P}_k=\{\set{P}_j\}_{j=1}^r$ satisfies the following properties.
  \begin{enumerate}
  \item For all $i\in\Naturals_k$, $\bigcard{\set{J}^{(i)}}\geq t-1$, and
  \item for all $j\neq j'\in\Naturals_r$, $\bigcard{\set{P}_j\cap\set{P}_{j'}}\leq 1$.
  \end{enumerate}
  Then, the corresponding systematic code $\code{C}(\collect{P}_k)$ is a $t$-server PIR code.
\end{lemma}

The above lemma only leads to an absorbing upper bound on the redundancy $N_{\textnormal{P}}(k,t)-k$ for fixed $t$ and
sufficiently large $k$, which shows that it is equal to $O(\sqrt{k})$. However, for smaller values of the parameter $k$,
whether or not this upper bound is tight is still unknown. Moreover, in \cite{FazeilVardyYaakobi15_1sub} a similar PIR
code construction based on constant-weight codes was provided, where all rows of $\mat{P}_{k\times r}$ have constant
weight and a given minimum Hamming distance.

It is known that the minimum Hamming distance $d$ of a PIR code must be larger than or equal to the desired parameter
$t$ (see Proposition~\ref{prop:d_tPIRcodes}), and so are the row Hamming weights of any generator matrix $\mat{G}$ for
the code. Hence, it is reasonable to change the sufficient condition of $\bigcard{\set{J}^{(i)}}\geq t-1$ in Lemma
\ref{lem:tPIRcode_Steiner} to $\bigcard{\set{J}^{(i)}}= t-1$, $\forall\,i\in\Naturals_k$.

Motivated by Steiner systems, we define a more elaborate systematic code family as follows.
\begin{definition}
  \label{def:code_propS}
  For any integer $t\in\Naturals$ and a given collection $\collect{P}_k=\{\set{P}_j\}_{j=1}^r$ of subsets of
  $\Naturals_k$, we say that a systematic code $\code{C}(\collect{P}_k)$ (or its corresponding generator matrix) has
  property $\prop{S}_t$ if all of the following conditions are satisfied.
  \begin{enumerate}[label={\arabic*)}]
  \item \label{item:1} $\set{P}_r=\Naturals_k$,
  \item \label{item:2} $\bigcard{\set{J}^{(i)}}= t-1$ for all $i\in\Naturals_k$,\smallskip
  \item \label{item:3} $\bigcard{\set{P}_j\cap\set{P}_{j'}}\leq 1$ for all $j\neq j'\in\Naturals_{r-1}$, and
  \item \label{item:4} for any given $m\in\Naturals_k$, there exists a subset $\set{I}(m)\subseteq\Naturals_k$ with
    $\set{I}(m)\cap\bigl(\bigcup_{j\in\set{J}^{(m)}\setminus\{r\}} \set{P}_j\bigr)=\emptyset$ and a subset
    $\set{V}^{(m)}\subseteq\Naturals_{r-1}$ with $\set{V}^{(m)}\cap\set{J}^{(m)}=\emptyset$ such that 
    \begin{IEEEeqnarray}{rCl}
      u_m+\sum_{i\in\set{I}(m)}u_i+\sum_{j\in\set{V}^{(m)}}\sum_{i\in\set{P}_j}u_i=\sum_{i=1}^k u_i.\notag
    \end{IEEEeqnarray}    
  \end{enumerate}
\end{definition}

Similarly to Lemma \ref{lem:tPIRcode_Steiner}, a systematic code with property $\prop{S}_t$ turns out to be an $[n,k;t]$
PIR code.
\begin{lemma}
  \label{lem:tPIRcode_propS}
  If a systematic code $\code{C}(\collect{P}_k)$ has property $\prop{S}_t$, then it is an $[n=k+r,k;t]$ PIR code.
\end{lemma}
\begin{IEEEproof}
  From conditions~\ref{item:2} and~\ref{item:3}, we know that for any $m\in\Naturals_k$,
  Lemma~\ref{lem:tPIRcode_Steiner} ensures that the code $\code{C}(\collect{P}_k\setminus\set{P}_r)$ has $t-1$ disjoint
  recovering sets for $u_m$. These mutually disjoint recovering sets can be expressed as follows:
  \begin{IEEEeqnarray*}{rCl}
    \set{R}^{(m)}_1& \eqdef &\{m\},
    \\
    \set{R}^{(m)}_{h+1}& \eqdef &\{\ell\in\set{P}_{j_h}\colon\ell\neq m\}\cup\{k+j_h\}, \quad h\in\Naturals_{t-2},
  \end{IEEEeqnarray*}
  where $j_h\in\set{J}^{(m)}\setminus\{r\}\eqdef\{j_1,\ldots,j_{t-2}\}$. 

  Moreover, the generator matrix of $\code{C}(\collect{P}_k)$ can be seen as a matrix that has an all-one column vector
  in the last column (with respect to $\set{P}_r=\Naturals_k$). Condition~\ref{item:4} directly implies that we can have
  one more recovering set defined by
  \begin{IEEEeqnarray*}{rCl}
    \set{R}^{(m)}_t\eqdef\set{I}(m)\cup\Bigl(\bigcup_{j\in\set{V}^{(m)}}\{k+j\}\Bigr)\cup\{k+r\},
  \end{IEEEeqnarray*}
  which is also disjoint from all other recovering sets by definition.
\end{IEEEproof}

The following example illustrates the code design of Lemma~\ref{lem:tPIRcode_propS}.
\begin{example}
  \label{ex:t5_k8n18}
  For an $[n,k]=[17,8]$ systematic code, we describe it in terms of $\collect{P}_8$ as follows:
  \begin{IEEEeqnarray*}{rCl}
    \collect{P}_8\eqdef\bigl\{\set{P}_1& \eqdef &\{1,2,3\},\set{P}_2\eqdef\{1,4,6\}, \set{P}_3\eqdef\{1,5,7\},
    \\
    \set{P}_4& \eqdef &\{2,4,8\},\set{P}_5\eqdef\{2,5,6\},\set{P}_6\eqdef\{3,4,7\},
    \\
    \set{P}_7& \eqdef &\{3,5,8\},\set{P}_8\eqdef\{6,7,8\},\set{P}_9\eqdef\Naturals_8\bigr\}.
  \end{IEEEeqnarray*}
  One can see that $r=9$ and that the systematic code $\code{C}(\collect{P}_8)$ has property $\prop{S}_5$. Here,
  condition \ref{item:4} can be verified by the following observations:
  \begin{IEEEeqnarray*}{rCl}
    \set{J}^{(1)}& = &\{1,2,3,9\},\;\set{J}^{(8)}=\{4,7,8,9\},
    \\
    \set{I}(1)& = &\{8\},\;\set{I}(8)=\{1\},\;\set{V}^{(1)}=\set{V}^{(8)}=\{5,6\},
    \\
    \Naturals_8& = &\{1\}\cup\set{P}_5\cup\set{P}_6\cup\{8\};
    \\
    \set{J}^{(2)}& = &\{1,4,5,9\},\;\set{J}^{(7)}=\{3,6,8,9\},
    \\
    \set{I}(2)& = &\{7\},\;\set{I}(7)=\{2\},\;\set{V}^{(2)}=\set{V}^{(7)}=\{2,7\},
    \\
    \Naturals_8& = &\{2\}\cup\set{P}_2\cup\set{P}_7\cup\{7\};
    \\
    \set{J}^{(3)}& = &\{1,6,7,9\},\;\set{J}^{(6)}=\{2,5,8,9\},
    \\
    \set{I}(3)& = &\{6\},\;\set{I}(6)=\{3\},\;\set{V}^{(3)}=\set{V}^{(6)}=\{3,4\},
    \\
    \Naturals_8& = &\{3\}\cup\set{P}_3\cup\set{P}_4\cup\{6\};
    \\
    \set{J}^{(4)}& = &\{2,4,6,9\},\;\set{J}^{(5)}=\{3,5,7,9\},
    \\
    \set{I}(4)& = &\{5\},\;\set{I}(5)=\{4\},\set{V}^{(4)}=\set{V}^{(5)}=\{1,8\},
    \\
    \Naturals_8& = &\{4\}\cup\set{P}_1\cup\set{P}_8\cup\{5\}.
  \end{IEEEeqnarray*}
  Then, we can conclude that this code is a $5$-server $[17,8]$ PIR code. For example, the recovering sets for the first
  information bit are determined by $\set{R}^{(1)}_1=\{1\}$,
  \begin{IEEEeqnarray*}{rCl}
    \set{R}_{2}^{(1)}& = &\{m\in\set{P}_1\colon m\neq
    1\}\cup\{k+1\}=\{2,3,9\},
    \\
    \set{R}_{3}^{(1)}& = &\{m\in\set{P}_2\colon m\neq
    1\}\cup\{k+2\}=\{4,6,10\},
    \\
    \set{R}_{4}^{(1)}& = &\{m\in\set{P}_3\colon m\neq
    1\}\cup\{k+3\}=\{5,7,11\},
    \\
    \set{R}_{5}^{(1)}& = &\{8,k+5,k+6,k+r\}=\{8,13,14,17\}.
  \end{IEEEeqnarray*}
\end{example}

In fact, the idea behind Lemma~\ref{lem:tPIRcode_propS} is to try to combine the properties of Steiner systems and
part~\ref{item:e} of Lemma~\ref{lem:bounds_tPIRcodes}, in such a way that we can construct an $[n+1,k;t+1]$ PIR code
from an $[n,k;t]$ PIR code when $t$ is even.

We also remark that a systematic $[n,k;t]$ PIR code with property $\prop{S}_t$ usually has different cardinalities of
its recovering sets (the so-called \emph{non-uniform information-symbol locality} property). For instance, for the code
of Example~\ref{ex:t5_k8n18}, each information symbol has $1$ recovering set of cardinality $1$, $3$ recovering sets of
cardinality $3$, and $1$ recovering set of cardinality $4$. This is also in alignment with
\cite{VajhaRamkumarKumar17_1}, where the presented PIR codes in general have recovering sets of different
cardinalities. In Section~\ref{sec:numerical-results}, we will show that codes having property $\prop{S}_5$ are good
$5$-server PIR codes with small blocklength.

\section{Searching for Optimal PIR Codes}
\label{sec:search_best-tPIRcodes}

In this section, we present an algorithm to search for good (i.e., small blocklength) PIR codes. Since optimal codes for
$t\leq 4$ are already known for all code dimensions $k$, we concentrate on $t=5$. Because
Theorem~\ref{thm:lengthen-tPIRcode} implies that we can construct a $t$-server PIR code by lengthening and extension,
hence, combined with the idea of lexicographic code construction \cite{Levenstein60_1},
Algorithm~\ref{alg:best-5PIRcodes} is proposed to find a sequence of good systematic PIR codes for $t=5$.\footnote{In
  general, this algorithm can be applied for any $t$. The main reason why we focus on small values of $t$ is that when
  $t$ is increasing, the complexity to determine whether a code has the $t$-PIR property is also increasing.}

\setlength{\textfloatsep}{14pt}
\begin{algorithm}[htbp]
  \renewcommand\AlCapFnt{\normalfont}
  \SetCommentSty{small}
  \caption{Searching for optimal $5$-server PIR codes}\label{alg:best-5PIRcodes}
  \DontPrintSemicolon
  \SetKwFunction{LengExt}{LengtheningExtending}
  \SetKwFunction{Lexi}{Lexical}
  \SetKwInOut{Input}{Input}
  \SetKwInOut{Output}{Output}
  
  \Input{A systematic constant row-weight-$5$ generator matrix
    $\mat{G}=[\mat{I}_k|\mat{P}_{k\times r}]$ for an $[n,k;5]$ code, and a given $w\in\Naturals_2$.}
  
  \Output{A systematic constant row-weight-$5$ generator matrix
    $\mat{G}_\textnormal{best}$ for an
    $[n_\textnormal{best},k_\textnormal{best};5]$ code, where
    $k_\textnormal{best} \geq k$ is the largest possible code
    dimension found and
    $n_\textnormal{best}=k_\textnormal{best}+r+w$.}
  
  $\mat{G}_\textnormal{best}\leftarrow [\mat{I}_k|\mat{P}_{k\times
    r}|\mat{O}_{k\times w}]$, $k_\textnormal{best}\leftarrow k$\;\label{alg-line:intl-Gbest}
  \tcc{$\mat{O}_{k\times w}$ is a $k \times w$ all-zero matrix}
  $i\leftarrow 1$\; $\vectg{z}\leftarrow$ the row vector
  $(1,1,1,1,0,\ldots,0)$ of length $r+w$\;\label{alg-line:z-vector}

  \While{$i\leq \binom{r+w}{4}$}{
    $\tilde{\mat{G}}\leftarrow$
    \LengExt{$\mat{G}_\textnormal{best},\vectg{z}$}\;\label{alg-line:LengExtG}
    $\tilde{d}\leftarrow$ minimum Hamming distance of $\tilde{\mat{G}}$\;
    \tcc{we simply say a code $\tilde{\mat{G}}$ is
      the set of all rows of $\tilde{\mat{G}}$}
    \If{$\tilde{d}\geq 6$\label{alg-line:distG}}{
      \eIf{$\tilde{\mat{G}}$ has the $5$-PIR
        property\label{alg-line:propS5}}{
        $\mat{G}_\textnormal{best} \leftarrow \tilde{\mat{G}}$,
        $k_\textnormal{best} \leftarrow k_\textnormal{best}+1$\;}
      {\KwRet{$(\mat{G}_\textnormal{best}, k_\textnormal{best})$}\;}}
    $i\leftarrow i+1,\vectg{z}\leftarrow$ \Lexi{$\vectg{z}$}\;}
  \If{$k_\textnormal{best}= k$}{$\mat{G}_\textnormal{best}\leftarrow \mat{G}$\;}
  \KwRet{$(\mat{G}_\textnormal{best}$, $k_\textnormal{best})$\;}
\end{algorithm}

Initially, we choose the best known $[n,k;5]$ code with a systematic generator matrix in which all rows have weight
$5$. Note that for small values of $n$ and $k$, such a code is not too difficult to find. As an example, the generator
matrix $\mat{G}$ of a systematic $[8,2;5]$ code in which all rows have weight $5$ is
\begin{IEEEeqnarray}{rCl}
  \mat{G} = \begin{bmatrix}[cc|ccccc|c]
    1 & 0 &  1 & 1 & 1 & 0 & 0 & 1
    \\
    0 & 1 &  1 & 0 & 0 & 1 & 1 & 1
  \end{bmatrix}.\label{eq:n8k2_t5}
\end{IEEEeqnarray}

The outer while loop of Algorithm~\ref{alg:best-5PIRcodes} increases a counter (denoted by $i$) from $1$ to
$\binom{r+w}{4}$ (the counter runs over all possible length-$(r+w)$ binary vectors of weight $4$). The function
$\texttt{LengtheningExtending}(\mat{G}_\textnormal{best},\vectg{z})$ in Line~\ref{alg-line:LengExtG} of
Algorithm~\ref{alg:best-5PIRcodes} is defined by
\begin{IEEEeqnarray*}{rCl}
  \tilde{\mat{G}}\eqdef
  \begin{bmatrix}[c|c|c]
    \mat{I}_{k_\textnormal{best}} & \vectg{0}&\mat{P}_{k_\textnormal{best} \times (r+w)}  
    \\*\cmidrule(lr){1-3}
    \tikzmark{n1}\vectg{0} & 1\tikzmark{n2}\tikzmark{n3}&\vectg{z}\tikzmark{n4} 
  \end{bmatrix},
  \tikz[overlay,remember picture]{
    \tikzunderbrace[raise=2mm,amplitude=4pt][yshift=-3mm]{$(n1)+(-0.3em,0em)$}{$(n2)+(0.4em,0em)$}{$k_\textnormal{best}+1$}
    \tikzunderbrace[raise=2mm,amplitude=4pt][yshift=-3mm]{$(n3)+(0.6em,0em)$}{$(n4)+(2.5em,0em)$}{$r+w$}
  }
\end{IEEEeqnarray*}
\bigskip

\noindent
where $\mat{G}_\textnormal{best}=\bigl[\mat{I}_{k_\textnormal{best}} |\mat{P}_{k_\textnormal{best}\times (r+w)}\bigr]$
and $w_\textnormal{H}(\vectg{z})=4$.\footnote{Note that the definition of $\tilde{\mat{G}}$ guarantees that
  $\mat{G}_\textnormal{best}$ is always in systematic form in each iteration.} Note that if $w=2$, it follows from the
proof of Theorem~\ref{thm:lengthen-tPIRcode} that $k_\textnormal{best} \geq k+1$; explaining why we choose
$1\leq w\leq 2$ from the beginning. Furthermore, notice that for $w=1$, sometimes the algorithm only results in the
original input code. We also verify whether $\tilde{d}\geq 6$ or not in Line~\ref{alg-line:distG} of
Algorithm~\ref{alg:best-5PIRcodes}. This is to ensure that the resulting code generated by $\tilde{\mat{G}}$ can
potentially satisfy Proposition~\ref{prop:d_tPIRcodes}.\footnote{Since the construction guarantees that all rows have
  equal Hamming weights, the Hamming distance between any pair of rows is even, i.e., the necessary condition
  $\tilde{d}\geq 5$ is equivalent to $\tilde{d}\geq 6$.}  Finally, given a vector $\vectg{z}$,
$\texttt{Lexical}(\vectg{z})$ generates the next lexicographical constant-weight $\vectg{z}$ of length $r+w$, e.g.,
$\texttt{Lexical}(\vectg{z})=(1,1,1,0,1,0,\ldots,0)$ for $\vectg{z}=(1,1,1,1,0,\ldots,0)$.

We also remark that the resulting $k_\textnormal{best}$ from Algorithm~\ref{alg:best-5PIRcodes} strongly depends on the
selected $\mat{G}=[\mat{I}_k|\mat{P}_{k\times r}]$ and the given $w$ in the input. It is difficult to predict whether
the corresponding blocklength $n_\textnormal{best}$ is good or not. For example, given the systematic
$[n=k+r,k;t]=[8,2;5]$ code defined in~\eqref{eq:n8k2_t5} and $w=1$, the output from Algorithm~\ref{alg:best-5PIRcodes}
is an $[n_\textnormal{best},k_\textnormal{best};5]=[11,4;5]$ code without property $\prop{S}_5$, while for $w=2$,
Algorithm~\ref{alg:best-5PIRcodes} results in an $[n_\textnormal{best},k_\textnormal{best};5]=[13,5;5]$ code with
property $\prop{S}_5$ (see Section~\ref{sec:numerical-results} that follows). Now, for code dimension $k=4$, the
$[11,4;5]$ code is better than the $[12,4;5]$ code obtained by shortening the optimal $[13,5;5]$ code. Hence, for a
fixed code dimension $k$, to find a good $5$-server PIR code with small blocklength, we have to compare all the
resulting $[n,k;5]$ codes found by Algorithm~\ref{alg:best-5PIRcodes}.

In general, the complexity of exhaustively examining the $t$-PIR property for a given code becomes infeasible for large
$n$ and $k$, even for $t=5$. However, according to our numerical results, for small code dimensions $k$, an optimal
$5$-server PIR code often has property $\prop{S}_5$. Therefore, we investigate a sequence of good PIR codes with respect
to property $\mat{S}_5$. In fact, a sequence of good codes with small blocklength can always be generated by lengthening
by one information symbol and extending at most $2$ coordinates from a smaller-sized code with property $\prop{S}_5$, as
shown in the theorem below.

\begin{theorem}
  \label{thm:lengthen-Scode}
  For any given values of $n$ and $k$, if a systematic $[n,k]$ code has property $\prop{S}_5$, then there must exist a
  systematic $[n+3,k+1]$ code that also has property $\prop{S}_5$.
\end{theorem}
\begin{IEEEproof}
  The proof idea is similar to the proof of Theorem~\ref{thm:lengthen-tPIRcode}, but involves a more intricate
  notation. The details are deferred to Appendix~\ref{sec:proof_lengthen-Scode} for better readability.
\end{IEEEproof}

Based on Theorem~\ref{thm:lengthen-Scode}, we can slightly modify Algorithm~\ref{alg:best-5PIRcodes} to investigate
$5$-server PIR codes with property $\prop{S}_5$. First, we replace the input generator matrix by a generator matrix
$\mat{G}=[\mat{I}_k|\mat{P}_{k\times (r-1)}|\vectg{1}]$ with property $\prop{S}_5$, and modify the starting
$\mat{G}_\textnormal{best}$ to $[\mat{I}_k|\mat{P}_{k\times (r-1)}|\mat{O}_{k\times w}|\vectg{1}]$ in
Line~\ref{alg-line:intl-Gbest} of Algorithm~\ref{alg:best-5PIRcodes}. The function
$\texttt{LengtheningExtending}(\mat{G}_\textnormal{best},\vectg{z})$ for
$\mat{G}_\textnormal{best}=\bigl[\mat{I}_{k_\textnormal{best}}|\mat{P}_{k_\textnormal{best}\times
  (r+w-1)}|\vectg{1}\bigr]$ in Line~\ref{alg-line:LengExtG} of Algorithm \ref{alg:best-5PIRcodes} is accordingly
re-defined as
\begin{IEEEeqnarray*}{rCl}
  \tilde{\mat{G}}\eqdef
  \begin{bmatrix}[c|c|c|c]
    \mat{I}_{k_\textnormal{best}} & \vectg{0}&\mat{P}_{k_\textnormal{best}\times (r+w-1)} & \vectg{1}
    \\*\cmidrule(lr){1-4} \tikzmark{n1}\vectg{0} & 1\tikzmark{n2}\tikzmark{n3}&\vectg{z} &\tikzmark{n4} 1
  \end{bmatrix},
  \tikz[overlay,remember picture]{
    \tikzunderbrace[raise=2mm,amplitude=4pt][yshift=-3mm]{$(n1)+(-0.3em,0em)$}{$(n2)+(0.4em,0em)$}{$k_\textnormal{best}+1$}
    \tikzunderbrace[raise=2mm,amplitude=4pt][yshift=-3mm]{$(n3)+(0.6em,0em)$}{$(n4)+(-0.6em,0em)$}{$r+w-1$}
  }
\end{IEEEeqnarray*}
\bigskip

\noindent
where $w_\textnormal{H}(\vectg{z})=5-2=3$. Notice that the outer while loop counter now should increase from $1$ to
$\binom{r+w-1}{3}$, and the initial $\vectg{z}$ in Line~\ref{alg-line:z-vector} should be replaced by the
length-$(r+w-1)$ vector $\vectg{z}=(1,1,1,0,\ldots,0)$. In fact, there is no need to modify Line~\ref{alg-line:distG} of
Algorithm~\ref{alg:best-5PIRcodes}, since the resulting $\tilde{\mat{G}}$ will again satisfy conditions
\ref{item:1}--\ref{item:3} of Definition~\ref{def:code_propS}.\footnote{Note that the construction of $\tilde{\mat{G}}$
  will make all the row-weights of $\tilde{\mat{G}}$ equal to $5$ and the last column equal to the all-one vector (i.e.,
  conditions \ref{item:1} and \ref{item:2} of Definition~\ref{def:code_propS} are satisfied). In order to satisfy
  condition~\ref{item:3} of Definition~\ref{def:code_propS}, the minimum Hamming distance of $\tilde{\mat{G}}$ must be
  larger than or equal to $2\cdot(5-2)=6$, since any two row vectors in $\tilde{\mat{G}}$ must have a common $1$ in at
  most two coordinates.} As a result, after the modifications to Algorithm \ref{alg:best-5PIRcodes} outlined above, and
if Line~\ref{alg-line:propS5} of Algorithm~\ref{alg:best-5PIRcodes} is replaced by the verification of property
$\prop{S}_5$ for $\tilde{\mat{G}}$, we are able to find good $5$-server PIR codes with property $\prop{S}_5$ for large
code dimensions $k\geq 16$ (see Section~\ref{sec:numerical-results} below). From Theorem~\ref{thm:lengthen-Scode} it
follows that if $w=2$, $k_\textnormal{best}\geq k+1$.

\section{Numerical Results}
\label{sec:numerical-results}

In this section, upper bounds on $N_\textnormal{P}(k,t)$ for $1\leq k\leq 32$ and $t=4,6,8$ are summarized in
Table~\ref{tab:table_6PIRcodes}. In particular, for $t=6$, we also present the numerical results obtained using the
search algorithm from Section~\ref{sec:search_best-tPIRcodes}. Entries for which strictly better codes are found than in
the current literature are marked in bold. In comparison with the obtained improved upper bound, a lower bound on
$N_\textnormal{P}(k,6)$ is also given. For $t=4$, the SPRM codes provided in \cite{VajhaRamkumarKumar17_1} are
optimal. More specifically, the blocklength is equal to the lower bound $L_\textnormal{P}(k,4)$.

In order to show how good our constructed $6$-server PIR codes are, we also list the best (smallest) known blocklength
for $t=8$ (the smallest blocklength of the SPRM codes from \cite{VajhaRamkumarKumar17_1}). They will result in an
improved upper bound for $t=6$, since by part~\ref{item:d} of Lemma~\ref{lem:bounds_tPIRcodes},
$N_\textnormal{P}(k,6)\leq N_\textnormal{P}(k,8)-2$. Hence,
\begin{IEEEeqnarray}{rCl}
  \label{eq:upperN_t6}
  n_\textnormal{U}\eqdef\min\{n_1,n_2-2\}
\end{IEEEeqnarray}
is the best known upper bound for $t=6$, where $n_1$ denotes the best known blocklength provided in
\cite{FazeilVardyYaakobi15_1sub}, and $n_2$ is the smallest blocklength of SPRM codes for $t=8$ provided in
\cite{VajhaRamkumarKumar17_1}.

Note again that, according to part~\ref{item:e} of Lemma~\ref{lem:bounds_tPIRcodes} and in order to compare our findings
with \cite[Table~III]{FazeilVardyYaakobi15_1sub} and \cite[Table~II]{VajhaRamkumarKumar17_1}, only even values of $t$
are interesting. Here, for $t=6$ the blocklengths $n_\textnormal{B}$ of Table~\ref{tab:table_6PIRcodes} are obtained by
adding one to the blocklengths of our best found $5$-server PIR codes. We make the following remarks to
Table~\ref{tab:table_6PIRcodes}.
\begin{table}[t]
  \centering
  \caption{Best known bounds on $N_\textnormal{P}(k,t)$ for small values of $k$ and even $t=4,6,8$. In the case of
    $t=6$, $n_\textnormal{B}$ denotes the best found blocklength based on our proposed search algorithm, and
    $n_\textnormal{U}$ is defined in \eqref{eq:upperN_t6}. Starred values (or columns) can be proved to be optimal,
    while bold entries are new results.}
  \label{tab:table_6PIRcodes}
  \vskip -2.0ex
  \begin{IEEEeqnarraybox}[
    \IEEEeqnarraystrutmode
    \IEEEeqnarraystrutsizeadd{0pt}{0pt}]{V/c/V/c/V/c/v/l/v/l/V/l/V}
    \IEEEeqnarrayrulerow\\
    & k \backslash t     
    && 4^{\ast\,\textnormal{\tiny\cite{VajhaRamkumarKumar17_1}}}
    && \IEEEeqnarraymulticol{5}{c}{$6$}
    && 8^{\textnormal{\tiny\cite{VajhaRamkumarKumar17_1}}}
    &\\    
    \hline\hline
    & 
    && 
    && N(k,t)^{\textnormal{\tiny\cite{Grassl:codetables}}} && n_\textnormal{B} && n_\textnormal{U}
    && 
    &\\    
    \hline\hline
    &1 && 4 &&- && 6^*&&-   &&8^* &\\
    \IEEEeqnarrayrulerow\\
    &2 && 6 &&- && 9^*&&-   &&12^*&\\
    \IEEEeqnarrayrulerow\\
    &3 && 7 &&- &&11^*&&-   &&14^*&\\
    \IEEEeqnarrayrulerow\\
    &4 && 9 &&- &&12^{*\, \diamond}&&-   &&15^*&\\
    \IEEEeqnarrayrulerow\\
    &5 &&10 &&14&&14^*&&13^!&&19  &\\
    \IEEEeqnarrayrulerow\\
    &6 &&11 &&15&&15^{*\,\diamond}&&14^!&&21  &\\
    \IEEEeqnarrayrulerow\\
    &7 &&13 &&16&&\mathbf{17}  &&15^!&&22  &\\
    \IEEEeqnarrayrulerow\\
    &8 &&14 &&17&&\mathbf{18}  &&20  &&24  &\\
    \IEEEeqnarrayrulerow\\
    &9 &&15 &&18&&\mathbf{20}  &&23  &&25  &\\
    \IEEEeqnarrayrulerow\\
    &10&&16 &&20&&\mathbf{21}  &&24  &&26  &\\
    \IEEEeqnarrayrulerow\\
    &11&&18 &&21&&\mathbf{22}  &&25  &&30  &\\
    \IEEEeqnarrayrulerow\\
    &12&&19 &&22&&\mathbf{23}&&26  &&32  &\\
    \IEEEeqnarrayrulerow\\
    &13&&20 &&23&&\mathbf{25}^\diamond&&27  &&33  &\\
    \IEEEeqnarrayrulerow\\
    &14&&21 &&24&&\mathbf{27}^\diamond&&29  &&35  &\\
    \IEEEeqnarrayrulerow\\
    &15&&22 &&26&&\mathbf{28}^\diamond&&34  &&36  &\\
    \IEEEeqnarrayrulerow\\
    &16&&24 &&27&&\mathbf{31}  &&35  &&37  &\\
    \IEEEeqnarrayrulerow\\
    &17&&25 &&28&&\mathbf{32}  &&37  &&39  &\\
    \IEEEeqnarrayrulerow\\
    &18&&26 &&29&&\mathbf{33}  &&38  &&40  &\\
    \IEEEeqnarrayrulerow\\
    &19&&27 &&30&&\mathbf{35}  &&39  &&41  &\\
    \IEEEeqnarrayrulerow\\
    &20&&28 &&31&&\mathbf{36}  &&40  &&42  &\\
    \IEEEeqnarrayrulerow\\
    &21&&29 &&32&&\mathbf{37}  &&42  &&46  &\\
    \IEEEeqnarrayrulerow\\
    &22&&31 &&33&&\mathbf{39}  &&46  &&48  &\\
    \IEEEeqnarrayrulerow\\
    &23&&32 &&34&&\mathbf{40}  &&47  &&49  &\\
    \IEEEeqnarrayrulerow\\
    &24&&33 &&36&&\mathbf{41}  &&49  &&51  &\\
    \IEEEeqnarrayrulerow\\
    &25&&34 &&37&&\mathbf{42}  &&50  &&52  &\\
    \IEEEeqnarrayrulerow\\
    &26&&35 &&38&&\mathbf{43}  &&51  &&53  &\\
    \IEEEeqnarrayrulerow\\
    &27&&36 &&39&&\mathbf{44}  &&53  &&55  &\\
    \IEEEeqnarrayrulerow\\
    &28&&37 &&40&&\mathbf{46}  &&54  &&56  &\\
    \IEEEeqnarrayrulerow\\
    &29&&39 &&41&&\mathbf{47}  &&55  &&57  &\\
    \IEEEeqnarrayrulerow\\
    &30&&40 &&42&&\mathbf{48}  &&56  &&58  &\\
    \IEEEeqnarrayrulerow\\
    &31&&41 &&43&&\mathbf{50}  &&58  &&60  &\\
    \IEEEeqnarrayrulerow\\
    &32&&42 &&44&&\mathbf{52}  &&59  &&61  &\\
    \IEEEeqnarrayrulerow    
  \end{IEEEeqnarraybox}    
\end{table}
\begin{enumerate}
\item The superscript ``$\ast$'' indicates that the corresponding blocklength can be shown to be optimal. We use the
  lower bound $N(k,t)$, whose value can be obtained from \cite{Grassl:codetables}, since
  $L_\textnormal{P}(k,6)=L_\textnormal{P}(k,4)+2\leq N(k,6)$ and no tighter lower bound for $t=6$ is known.
  
\item The superscript ``$\diamond$'' indicates that the best found systematic $[n,k;5]$ code has a constant-weight
  generator matrix of row-weight $5$ and without property $\prop{S}_5$.

\item The superscript ``$!$'' indicates that the corresponding blocklength is impossible, since it is smaller than
  $N(k,t)$ (a contradiction to Corollary~\ref{cor:lower-bound_tPIRcodes}). We believe that the value of $n_\text{U}=15$
  for $(k,t)=(7,6)$ in \cite[Table~III]{FazeilVardyYaakobi15_1sub} was obtained from
  \cite[Thm.~9]{FazeilVardyYaakobi15_1sub} and should have corresponded to $(k,t)=(6,6)$ due to a misprint in
  \cite[p.~289]{lincostello04_1} in the redundancy of \emph{type-$1$ doubly transitive invariant codes} (see the
  algebraic code construction in \cite[Thm.~6]{LinMarkowsky90_1}). We believe this explains the contradictions.

\item The superscript ``[$\cdot$]'' indicates the reference number.
\end{enumerate}  

We also remark that for $t=8$, using our algorithm we are able to find better PIR codes for certain values of $k$: we
have obtained $n_\textnormal{B}=18,20,29,31$ for $k=5,6,11,12$, respectively. This indicates that the SPRM codes are not
optimal for $t=8$.



\section{Conclusion}
\label{sec:conclusion}

In this paper, we presented a construction of a $t$-server PIR code by lengthening and extension of an existing PIR
code. We also presented an extension of a code construction inspired by Steiner systems proposed by Fazeli \emph{et
  al.}, which was used in the proposed algorithm to search for good (i.e., small blocklength) $5$-server PIR codes. For
code dimensions $k \leq 6$, provably optimal PIR codes were found, while for all $7 \leq k \leq 32$, codes of smaller
blocklength than the best known codes from the literature were found and presented. Moreover, better $8$-server PIR
codes were also found for $k=5,6,11,12$.


\appendices


\section{Proof of Theorem~\ref{thm:lengthen-tPIRcode}}
\label{sec:proof_lengthen-tPIRcode}

We firstly consider the case of $t$ even. Assume that there exists an $[n,k;t]$ PIR code with even $t$ and generator
matrix denoted by $\mat{G}_{k \times n}=[\vectg{c}_1,\ldots,\vectg{c}_n]$. We will prove that there always exists an
$[n+t/2,k+1;t]$ code with generator matrix
\begin{IEEEeqnarray*}{rCl}
  \tilde{\mat{G}}& = &
  \begin{bmatrix}        
    \mat{G}_{k\times n}  &\mat{O}_{k\times \frac{t}{2}}
    \\
    \vectg{z}             &\tikzmark{n1}\quad\vectg{1}\quad\tikzmark{n2}
  \end{bmatrix}
  \\
  & \eqdef &
  \begin{bmatrix}
    \tilde{\vectg{c}}_1,\cdots,\tilde{\vectg{c}}_n,\tilde{\vectg{c}}_{n+1},\cdots,\tilde{\vectg{c}}_{n+t/2}
  \end{bmatrix},
\end{IEEEeqnarray*}
where $\vectg{z}$ is a length-$n$ binary vector of Hamming weight $w_\textnormal{H}(\vectg{z})=\frac{t}{2}$. In the
following, let $\vectg{e}^{(j)}_{i}$ denote the $i$-th unit vector of length $j$, from which it follows that
$\tilde{\vectg{c}}_{n+j}=\vectg{e}^{(k+1)}_{k+1}$, $\forall\,j\in\Naturals_{t/2}$.

To prove the existence, first, since we know that $\mat{G}_{k \times n}$ is the generator matrix of an $[n,k;t]$ code,
given an information index $m\in\Naturals_k$, there exists a collection of mutually disjoint recovering sets
$\collect{R}^{(m)}=\{\set{R}^{(m)}_1,\ldots,\set{R}^{(m)}_{t}\}$ for $u_m$. Let us arbitrarily choose $t/2$ recovering
sets $\set{R}^{(m)}_{h_s}\in\collect{R}^{(m)}$, $1\leq s\leq t/2$, and then select exactly one $j_s$ from each
$\set{R}^{(m)}_{h_s}$, $s\in\Naturals_{t/2}$. We then define a row vector $\vectg{z}$ with Hamming weight
$w_\textnormal{H}(\vectg{z})=t/2$ as
\begin{IEEEeqnarray*}{rCl}
  z_{j_s}=
  \begin{cases}
    1, & \textnormal{if } j_s\in\set{R}^{(m)}_{h_s},\;s\in\Naturals_{t/2},
    \\
    0, & \textnormal{otherwise}.
  \end{cases}
\end{IEEEeqnarray*}
Define
\begin{IEEEeqnarray*}{rCl}
  \tilde{\set{R}}^{(k+1)}_h\eqdef
  \begin{cases}
    \set{R}^{(m)}_{h_s}\cup\set{R}^{(m)}_{h'_s}, & \textnormal{for } h,s\in\Naturals_{t/2},
    \\
    \{n+h-t/2\}, & \textnormal{for } h\in\Naturals_{t}\setminus\Naturals_{t/2},
  \end{cases}
\end{IEEEeqnarray*}
where $\set{R}^{(m)}_{h'_s}$ are the recovering sets in $\collect{R}^{(m)}$ other than $\set{R}^{(m)}_{h_s}$,
$s\in\Naturals_{t/2}$. According to the definition of PIR codes, the sets $\tilde{\set{R}}^{(k+1)}_h$,
$h\in\Naturals_t$, must be disjoint. Furthermore, for every $h\in\Naturals_{t/2}$, we have
\begin{IEEEeqnarray}{rCl}
  \sum_{j\in\tilde{\set{R}}_h^{(k+1)}}\tilde{\vectg{c}}_j& = &
  \sum_{j\in\set{R}_{h_s}^{(m)}}\tilde{\vectg{c}}_j+\sum_{j\in\set{R}_{h'_s}^{(m)}}\tilde{\vectg{c}}_j
  \nonumber\\*
  & = &
  \begin{bmatrix}
    \vectg{e}^{(k)}_m
    \\
    1
  \end{bmatrix}
  +
  \begin{bmatrix}
    \vectg{e}^{(k)}_m
    \\
    0
  \end{bmatrix}=\vectg{e}^{(k+1)}_{k+1}, \label{eq:1}
\end{IEEEeqnarray}
where \eqref{eq:1} holds by our definition of $\vectg{z}$. Thus, for each $h\in\Naturals_{t/2}$,
$\tilde{\set{R}}^{(k+1)}_h$ is a recovering set. Together with the recovering sets
$\tilde{\set{R}}_h^{(k+1)}=\{n+h-t/2\}$, $h\in\Naturals_t\setminus\Naturals_{t/2}$, this guarantees that the extra
information bit $u_{k+1}$ has $2\cdot \frac{t}{2}=t$ disjoint recovering sets.

In the second part of the proof, we prove that the number of recovering sets for any information bit $u_i$,
$i\in\Naturals_k$, remains unchanged (i.e., it is equal to $t$). For any given message index $i\in\Naturals_k$, since
$w_\textnormal{H}(\vectg{z})=t/2$, there are at most $t/2$ recovering sets $\set{R}^{(i)}$ for which
$\bigcard{\set{R}^{(i)}\cap\{j_s\}_{s=1}^{t/2}}$ is odd, where $\{j_s\}_{s=1}^{t/2}$ are the selected column indices
from the first part of the proof. Without loss of generality, we can then reorder these recovering sets as
$\set{R}^{(i)}_1,\ldots,\set{R}^{(i)}_v$, for some integer $v\leq t/2$, such that for each $h\in\Naturals_v$,
$\bigcard{\set{R}^{(i)}_h\cap\{j_s\}_{s=1}^{t/2}}$ is odd.  Accordingly, for the remaining recovering sets
$\set{R}^{(i)}_{v+1},\ldots,\set{R}^{(i)}_{t}$, $\bigcard{\set{R}^{(i)}_h\cap\{j_s\}_{s=1}^{t/2}}$,
$h\in\Naturals_t\setminus\Naturals_v$, is even.

For each $h\in\Naturals_t$, the recovering sets for $u_i$ of the new $[n+t/2,k;t]$ code can then be defined as
\begin{IEEEeqnarray*}{rCl}
  \tilde{\set{R}}^{(i)}_h\eqdef
  \begin{cases}
    \set{R}^{(i)}_h\cup\{n+h\}, &h\in\Naturals_{v},
    \\
    \set{R}^{(i)}_h, & \textnormal{otherwise}.
  \end{cases}
\end{IEEEeqnarray*}
Hence,
\begin{IEEEeqnarray}{rCl}
  \sum_{j\in\tilde{\set{R}}_h^{(i)}}\tilde{\vectg{c}}_j
  & = &
  \begin{cases}
    \sum_{j\in\set{R}_h^{(i)}}\tilde{\vectg{c}}_j+\vectg{e}^{(k+1)}_{k+1},
    &h\in\Naturals_{v},
    \\[2mm]
    \sum_{j\in\set{R}_h^{(i)}}\tilde{\vectg{c}}_j, &
    \textnormal{otherwise}
  \end{cases}
  \nonumber\\[1mm]
  & = &
  \begin{cases}
    \begin{bmatrix}
      \vectg{e}^{(k)}_i
      \\
      1
    \end{bmatrix}+\vectg{e}^{(k+1)}_{k+1}=\vectg{e}^{(k+1)}_i,
    &h\in\Naturals_{v},
    \\[4mm]
    \vectg{e}^{(k+1)}_i, & \textnormal{otherwise},
    \label{eq:2}
  \end{cases}\IEEEeqnarraynumspace
\end{IEEEeqnarray}
where \eqref{eq:2} holds because we assume that for each $h\in\Naturals_v$,
$\bigcard{\set{R}^{(i)}_h\cap\{j_s\}_{s=1}^{t/2}}$ is odd. Hence, for $t$ even,
\begin{IEEEeqnarray*}{rCl}
  N_\textnormal{P}(k+1,t)& \leq &N_\textnormal{P}(k,t+1)+\frac{t}{2}.
\end{IEEEeqnarray*}

When $t$ is odd (and thus $t+1$ even), we have
\begin{IEEEeqnarray}{rCl}
  N_\textnormal{P}(k+1,t)+1& = &N_\textnormal{P}(k+1,t+1) \label{eq:a}\\
  & \leq & N_\textnormal{P}(k,t+1)+\frac{t+1}{2} \label{eq:b}
  \\
  & = &N_\textnormal{P}(k,t)+1+\frac{t+1}{2}, \label{eq:c}
\end{IEEEeqnarray}
where \eqref{eq:a} and \eqref{eq:c} follow from part~\ref{item:e} of Lemma~\ref{lem:bounds_tPIRcodes} and \eqref{eq:b}
from the result above, which is equivalent to
\begin{IEEEeqnarray*}{rCl}
  N_\textnormal{P}(k+1,t)\leq N_\textnormal{P}(k,t)+\left\lceil\frac{t}{2}\right\rceil.
\end{IEEEeqnarray*}

\section{Proof of Theorem~\ref{thm:lengthen-Scode}}
\label{sec:proof_lengthen-Scode}

Before we start the proof, recall that a systematic $[n=k+r,k]$ code is said to have property $\prop{S}_5$ if its
corresponding collection $\collect{P}_k$ satisfies all the conditions of Definition~\ref{def:code_propS}. Accordingly,
for each $m\in\Naturals_k$, we write the corresponding sets for collection
$\collect{P}_k=\{\set{P}_1,\ldots,\set{P}_r\}$ in Definition \ref{def:code_propS} as $\set{J}^{(m)}$, $\set{I}(m)$, and
$\set{V}(m)$. We also denote the generator matrix of the code as
$\mat{G}=[\mat{I}_k| \mat{P}_{k\times (r-1)} | \vectg{1}]$.

Next, we consider a systematic $[(k+1)+(r+2),k+1]$ code whose generator matrix is given by
\begin{IEEEeqnarray*}{rCl}
  \tilde{\mat{G}}\eqdef
  \begin{bmatrix}[c|c|c|cc|c]
    \mat{I}_k & \vectg{0}&\mat{P}_{k\times (r-1)}& \vectg{0} &\vectg{0} &
    \vectg{1}
    \\*\cmidrule(lr){1-6}
    \tikzmark{n1}\vectg{0} & 1\tikzmark{n2}\tikzmark{n3}&\vectg{z} &\tikzmark{n4} 1 & 1 & 1
  \end{bmatrix},
  \tikz[overlay,remember picture]{
    \tikzunderbrace[raise=2mm,amplitude=4pt][yshift=-3mm]{$(n1)+(-0.3em,0em)$}{$(n2)+(0.4em,0em)$}{$k+1$}
    \tikzunderbrace[raise=2mm,amplitude=4pt][yshift=-3mm]{$(n3)+(0.6em,0em)$}{$(n4)+(-0.6em,0em)$}{$r-1$}
  }
\end{IEEEeqnarray*}
\bigskip

\noindent
where $w_\textnormal{H}(\vectg{z})=1$ (i.e., the Hamming weight of the last row of $\tilde{\mat{G}}$ is equal to
$5$). This will result in a new collection
\begin{IEEEeqnarray*}{rCl}
  \IEEEeqnarraymulticol{3}{l}{%
    \tilde{\collect{P}}_{k+1}\eqdef\{\set{P}_j\}_{j=1}^{\tilde{r}}
  }\nonumber\\*\quad%
  & = &
  \{\tilde{\set{P}}_1,\ldots,\tilde{\set{P}}_{r-1},\nonumber\\
  &&\;\>\tilde{\set{P}}_r=\{k+1\},\tilde{\set{P}}_{r+1}=\{k+1\},
  \tilde{\set{P}}_{r+2}=\Naturals_{k+1}\}.
\end{IEEEeqnarray*}
Notably, $\tilde{r}=r+2$ corresponds to the all-one column of $\tilde{\mat{G}}$. In addition, since
$w_\textnormal{H}(\vectg{z})=1$, we know that there exists only one $j_1\in\Naturals_{r-1}$ such that
$\tilde{\set{P}}_{j_1}=\set{P}_{j_1}\cup\{k+1\}$, and all other sets of $\{\tilde{\set{P}}_j\}_{j=1}^{r-1}$ are
unchanged, i.e., $\tilde{\set{P}}_j=\set{P}_j$, $\forall\;j\in\Naturals_{r-1}\setminus\{j_1\}$. Hence, it is trivial to
see that $\code{C}(\tilde{\collect{P}}_{k+1})$ satisfies conditions \ref{item:1}--\ref{item:3} of property $\mat{S}_5$.

Regarding condition \ref{item:4}, because we know how the new collection of $\tilde{\collect{P}}_{k+1}$ will be, we can
show that
\begin{IEEEeqnarray*}{rCl}
  \tilde{\set{J}}^{(m)}& = &\bigl(\set{J}^{(m)}\setminus\{r\}\bigr)\cup\{r+2\}, m\in\Naturals_k,
  \textnormal{ and}\\
  \tilde{\set{J}}^{(k+1)}& = &\{j_1,r,r+1,r+2\}
\end{IEEEeqnarray*}
with respect to the new code $\code{C}(\tilde{\collect{P}}_{k+1})$.

Our goal is to verify that for any $m\in\Naturals_{k+1}$, there exist two subsets
$\tilde{\set{I}}(m)\subseteq\Naturals_{k+1}$ with
$\tilde{\set{I}}(m)\cap\bigl(\bigcup_{j\in\tilde{\set{J}}^{(m)}\setminus\{r+2\}} \tilde{\set{P}}_j\bigr)=\emptyset$ and
$\tilde{\set{V}}^{(m)}\subseteq\Naturals_{r+1}$ with $\tilde{\set{V}}^{(m)}\cap\tilde{\set{J}}^{(m)}=\emptyset$, such
that
\begin{IEEEeqnarray}{rCl}
  u_m+\sum_{i\in\tilde{\set{I}}(m)}u_i+\sum_{j\in\tilde{\set{V}}^{(m)}}\sum_{i\in\tilde{\set{P}}_j}u_i
  =\sum_{i=1}^{k+1}u_i.\label{eq:check_propS_k+1}
\end{IEEEeqnarray}

First, consider the case when $m\in\Naturals_k$. Since $\code{C}(\collect{P}_k)$ has property $\prop{S}_5$, there exist
two sets $\set{I}(m)\subseteq\Naturals_k$ and $\set{V}^{(m)}\subseteq\Naturals_{r-1}$ that satisfy
condition~\ref{item:4}. Hence, one can check that \eqref{eq:check_propS_k+1} also holds if we define
\begin{IEEEeqnarray*}{rCl}
  \tilde{\set{I}}(m)& \eqdef &\set{I}(m),
  \textnormal{ and}
  \\
  \tilde{\set{V}}^{(m)}& \eqdef &
  \begin{cases}
    \set{V}^{(m)}, &\textnormal{if }j_1\in\set{V}^{(m)},
    \\
    \set{V}^{(m)}\cup\{r\}, &\textnormal{if }j_1\notin\set{V}^{(m)}.
  \end{cases}
\end{IEEEeqnarray*}
Note again that $\tilde{\set{J}}^{(m)}\setminus\{r+2\}=\set{J}^{(m)}\setminus\{r\}$,
$\set{I}(m)=\tilde{\set{I}}(m)\subseteq\Naturals_k$, $\set{V}^{(m)}\subseteq\Naturals_{r-1}$, and
$\tilde{\set{V}}^{(m)}\subseteq\Naturals_r$. We substantiate the disjoint requirements for $m\in\Naturals_k$ by
observing that
\begin{IEEEeqnarray*}{rCl}
  \IEEEeqnarraymulticol{3}{l}{%
    \tilde{\set{I}}(m)\cap\bigl(\bigcup_{j\in\tilde{\set{J}}^{(m)}\setminus\{r+2\}}\tilde{\set{P}}_j\bigr)
    =\set{I}(m)\cap\bigl(\bigcup_{j\in\set{J}^{(m)}\setminus\{r\}}\tilde{\set{P}}_j\bigr)}\nonumber\\*\;%
  \\
  & = &
  \begin{cases}
    \set{I}(m)\cap\Bigl[\bigl(\bigcup_{j\in\set{J}^{(m)}\setminus\{r\}}\set{P}_j\bigr)\cup\{k+1\}\Bigr],
    &\textnormal{if } j_1\in\set{J}^{(m)},
    \\[2mm]
    \set{I}(m)\cap\bigl(\bigcup_{j\in\set{J}^{(m)}\setminus\{r\}}\set{P}_j\bigr),
    & \textnormal{otherwise}
  \end{cases}
  \\[1mm]
  & = &\emptyset,
\end{IEEEeqnarray*}
and
\begin{IEEEeqnarray*}{rCl}
  \tilde{\set{V}}^{(m)}\cap\tilde{\set{J}}^{(m)}& = &
  \tilde{\set{V}}^{(m)}\cap\bigl(\tilde{\set{J}}^{(m)}\setminus\{r+2\}\bigr)
  \\
  & = &\tilde{\set{V}}^{(m)}\cap(\set{J}^{(m)}\setminus\{r\}\bigr)
  \\
  & = &\set{V}^{(m)}\cap(\set{J}^{(m)}\setminus\{r\}\bigr)=\emptyset.
\end{IEEEeqnarray*}

Secondly, for the newly added information index $k+1$, choose one $\ell\in\tilde{\set{P}}_{j_1}$ such that
$\ell\neq k+1$. Since we know that \eqref{eq:check_propS_k+1} holds for $\ell\in\Naturals_k$, we have
\begin{IEEEeqnarray}{rCl}
  u_\ell+ \sum_{i\in\tilde{\set{I}}(\ell)}u_i+
  \sum_{j\in\tilde{\set{V}}^{(\ell)}}\sum_{i\in\tilde{\set{P}}_j}u_i
  =\sum_{i=1}^{k+1}u_i \label{eq:check_ul}
\end{IEEEeqnarray}
and $\tilde{\set{J}}^{(\ell)}\setminus\{r+2\}\eqdef\{j_1,j_2,j_3\}=\set{J}^{(\ell)}\setminus\{r\}$ for some
$j_2\neq j_3\in\Naturals_{r-1}$. Noting that since $\set{V}^{(\ell)}\cap\set{J}^{(\ell)}=\emptyset$ and
$j_1\notin\set{V}^{(\ell)}$, we have $\tilde{\set{V}}^{(\ell)}=\set{V}^{(\ell)}\cup\{r\}$. Therefore, assume that
\begin{IEEEeqnarray}{rCl}
  \tilde{\set{I}}(k+1)& \eqdef &\tilde{\set{I}}(\ell)\cup\bigl(\tilde{\set{P}}_{j_2}\setminus\{\ell\}\bigr), \text{ and}
  \nonumber\\[1mm]
  \tilde{\set{V}}^{(k+1)}& \eqdef &\bigl(\tilde{\set{V}}^{(\ell)}\setminus\{r\}\bigr)\cup\{j_2\}
  =\set{V}^{(\ell)}\cup\{j_2\}.\label{eq:sets_u_new}
\end{IEEEeqnarray}
Then, the disjoint requirements are satisfied by confirming that
\begin{IEEEeqnarray}{rCl}
  \IEEEeqnarraymulticol{3}{l}{%
    \tilde{\set{I}}(k+1)\cap\Bigl(\bigcup_{j\in\tilde{\set{J}}^{(k+1)}\setminus\{r+2\}}\tilde{\set{P}}_j\Bigr)
  }\nonumber\\*\quad%
  & = &\bigl[\tilde{\set{I}}(\ell)\cup(\tilde{\set{P}}_{j_2}\setminus\{\ell\})\bigr]
  \cap\bigl(\tilde{\set{P}}_{j_1}\cup\{k+1\}\cup\{k+1\}\bigr)
  \IEEEeqnarraynumspace\nonumber\\
  & = &\bigl[\tilde{\set{I}}(\ell)\cup(\tilde{\set{P}}_{j_2}\setminus\{\ell\})\bigr]
  \cap\tilde{\set{P}}_{j_1}\IEEEeqnarraynumspace\label{eq:3}
  \\
  & = &\bigl(\tilde{\set{I}}(\ell)\cap\tilde{\set{P}}_{j_1}\bigr)\cup
  \bigl[(\tilde{\set{P}}_{j_2}\setminus\{\ell\})\cap\tilde{\set{P}}_{j_1}\bigr]
  \nonumber\\
  & = &\emptyset\cup \bigl[(\tilde{\set{P}}_{j_2}\setminus\{\ell\})\cap\tilde{\set{P}}_{j_1}\bigr]\label{eq:4}
  \\
  & = &\emptyset,\label{eq:5}
  \\[2mm]
  \IEEEeqnarraymulticol{3}{l}{%
    \tilde{\set{V}}^{(k+1)}\cap\bigl(\tilde{\set{J}}^{(k+1)}\setminus\{r+2\}\bigr)} \nonumber\\*\quad%
  & = &\bigl(\set{V}^{(\ell)}\cup\{j_2\}\bigr)\cap\{j_1,r,r+1\}
  \nonumber\\[1mm]
  & = &\bigl(\set{V}^{(\ell)}\cap\{j_1\}\bigr)\cup\emptyset=\emptyset.\IEEEeqnarraynumspace\label{eq:6}
\end{IEEEeqnarray}
Here, \eqref{eq:3} follows since $\tilde{\set{P}}_{j_1}=\set{P}_{j_1}\cup\{k+1\}$; in \eqref{eq:4}, we use the fact that
$\tilde{\set{I}}(\ell)\cap\bigl(\bigcup_{j\in\tilde{\set{J}}^{(\ell)}\setminus\{r+2\}}
\tilde{\set{P}}_j\bigr)=\emptyset$; \eqref{eq:5} follows because $\ell\in\tilde{\set{P}}_{j_1}\cap\tilde{\set{P}}_{j_2}$
and $\bigcard{\tilde{\set{P}}_{j_1}\cap\tilde{\set{P}}_{j_2}}\leq 1$; and \eqref{eq:6} holds by assumption since
$\set{V}^{(\ell)}$ and $\set{J}^{(\ell)}=\{j_1,j_2,j_3,r\}$ are disjoint.

Finally, from \eqref{eq:check_ul} we obtain
\begin{IEEEeqnarray}{rCl}
  \sum_{i=1}^{k+1}u_i& = &u_\ell+ \sum_{i\in\tilde{\set{I}}(\ell)}u_i+
  \sum_{j\in\tilde{\set{V}}^{(\ell)}}\sum_{i\in\tilde{\set{P}}_j}u_i
  \nonumber\\
  & =
  &\left(\sum_{i\in\tilde{\set{P}}_{j_2}}u_i+\sum_{i\in\tilde{\set{P}}_{j_2}\setminus\{\ell\}}u_i\right)\nonumber\\
  &&\>+\sum_{i\in\tilde{\set{I}}(\ell)}u_i+ \sum_{j\in\tilde{\set{V}}^{(\ell)}}\sum_{i\in\tilde{\set{P}}_j}u_i
  \nonumber\\
  & = &\sum_{i\in\tilde{\set{P}}_{j_2}\setminus\{\ell\}}u_i+
  \sum_{i\in\tilde{\set{I}}(\ell)}u_i+\sum_{i\in\tilde{P}_r}u_i
  \nonumber\\
  &&\>+\sum_{j\in\tilde{\set{V}}^{(\ell)}\setminus\{r\}}\sum_{i\in\tilde{\set{P}}_j}u_i
  +\sum_{i\in\tilde{\set{P}}_{j_2}}u_i
  \nonumber\\
  & = &u_{k+1}+\sum_{i\in\tilde{\set{I}}(k+1)}u_i+
  \sum_{j\in\tilde{\set{V}}^{(k+1)}}\sum_{i\in\tilde{\set{P}}_j}u_i,\label{eq:7}
\end{IEEEeqnarray}
where \eqref{eq:7} holds because of \eqref{eq:sets_u_new} and $\tilde{\set{P}}_{r}=\{k+1\}$. Thus, we have proved that
for all $m\in\Naturals_{k+1}$, there are explicit sets $\tilde{\set{I}}(m)$ and $\tilde{\set{J}}^{(m)}$ that validate
\eqref{eq:check_propS_k+1}.

\balance 

\bibliographystyle{IEEEtran}
\bibliography{defshort1,biblio1}

\end{document}